\documentclass[twocolumn]{IEEEtran} 

\def\BibTeX{{\rm B\kern-.05em{\sc i\kern-.025em b}\kern-.08em
    T\kern-.1667em\lower.7ex\hbox{E}\kern-.125emX}}

\newcommand\beq{\begin{equation}}
\newcommand\eeq{\end{equation}}
\newcommand\rhobf{\mbox{\boldmath $\rho$}}
\newcommand\kappabf{\mbox{\boldmath $\kappa$}}
\newcommand\alphabf{\mbox{\boldmath $\alpha$}}

\usepackage[final]{graphicx}
\usepackage{amsmath}
\usepackage{amsfonts}   
\usepackage{amssymb}
\begin{document}


\title{Short-Pulsed Wavepacket Propagation\protect\\
in Ray-Chaotic Enclosures}

\author{Giuseppe Castaldi, Vincenzo Galdi, {\em Senior Member, IEEE}, and Innocenzo M. Pinto, {\em Senior Member, IEEE}\thanks{
The authors are with the  Waves Group, Department of Engineering, University of Sannio, I-82100 Benevento, Italy (e-mail: castaldi@unisannio.it, vgaldi@unisannio.it, pinto@sa.infn.it).
}}

\maketitle

\begin{abstract}
Wave propagation in {\em ray-chaotic} scenarios, characterized by exponential sensitivity to ray-launching conditions, is a topic of significant interest, with deep phenomenological implications and important applications, ranging from optical components and devices to time-reversal focusing/sensing schemes. Against a background of available results that are largely focused on the time-harmonic regime, we deal here with {\em short-pulsed} wavepacket propagation in a ray-chaotic enclosure. For this regime, we propose a rigorous analytical framework based on a short-pulsed {\em random-plane-wave} statistical representation, and check its predictions against the results from finite-difference-time-domain numerical simulations. 
\end{abstract}

\markboth{CASTALDI {\em et al.}: Short-Pulsed Wavepacket Propagation in Ray-Chaotic Enclosures}{} 


\section{Introduction and Background}
\PARstart{D}{uring} the last decades, there has been a growing interest in the study of electromagnetic (EM) wave propagation in environments featuring {\em ray-chaotic} behavior, i.e., {\em exponential} separation of nearby-originating ray trajectories \cite{Lichtenberg,Ott}.
Remarkably, the onset of such odd behavior is not necessarily a mere consequence of geometrical/structural complexity, but may occur even in {\em deceptively simple} coordinate-non-separable geometries with {\em linear} constitutive relationships as an effect of the inherent nonlinearity of the ray equation \cite{Born}. By exploiting the formal analogy with ``billiards'' and ``pinballs'' from classical mechanics (see, e.g., \cite{Sinai}--\cite{Gaspard} for a sparse sampling, and \cite{Chernov} for a comprehensive review of the key concepts and ideas), simple examples of ray-chaotic closed (cavity-type) or open (scattering) propagation scenarios may be readily conceived (see, e.g., \cite{Smilansky}--\cite{Ramahi}). 

Besides the critical implications in the actual applicability of ray-based theory and related computational schemes (see, e.g., \cite{Mackay1}), ray-chaotic manifestations have been shown to play a key role in several fields of application, including 
optical microcavities and lasers \cite{Noeckel2}--\cite{Shinohara},
characterization of scattering signatures \cite{Mackay2,Mackay3},  time-reversal-based focusing \cite{Fink1,Anlage4} and sensing \cite{Anlage5}, and reverberation chambers \cite{Cappetta}--\cite{Orjubin2}.

From the phenomenological viewpoint, a large body of theoretical, numerical and experimental studies have evidenced the presence of {\em distinctive} features (often with {\em universal} character) in the time-harmonic high-frequency wave dynamics observed in ray-chaotic scenarios, in terms, e.g., of the ensemble properties of eigenvalues,  eigenfunctions, and impedance/scattering matrices, in close analogy with {\em quantum chaos} (see, e.g., \cite{Berry}--\cite{Stockmann} and references therein). These features
are not observable in ray-regular (e.g., coordinate-separable) configurations, and
are connected to the spectral properties exhibited by certain classes of random matrices \cite{RM}.
Hence, although the (linear) wave dynamics must be {\em non-chaotic} in strict technical sense, it does exhibit ``ray-chaotic footprints'' in the high-frequency regime. These, with a few notable exceptions (see, e.g., \cite{Heller1,Sridhar}), generally result into an irregular random-like behavior. In such scenario, ray-based techniques are deterministically inaccurate and full-wave predictions may not be necessary for certain applications, while {\em statistical} models based on random-plane-wave (RPW) superpositions \cite{Berry1,Heller2} may be more insightful and useful to capture and parameterize the essential physics. Interestingly, similar RPW models were independently developed within the framework of narrowband EM reverberation chambers \cite{Hill}.

Against the above background results, mostly focused on time-harmonic excitation (for which the quantum/EM analogy may be exploited to its fullest), this paper deals with {\em short-pulsed} (SP) wavepacket propagation in ray-chaotic enclosures. For this regime, comparatively fewer results are available in the quantum-physics literature (mostly dealing with ``Loschmidt echoes'' and ``fidelity decay'' \cite{Gorin}), and their translation to EM scenarios is less straightforward in view of the limited (to the paraxial regime) equivalence between the wave equation and the time-dependent Schr\"odinger equation. Among the EM results available, it is worth mentioning those pertaining to ``one-channel'' time-reversal focusing \cite{Anlage4} and related fidelity-based sensing schemes \cite{Anlage5}, where the {\em ergodicity} properties associated with ray chaos are exploited in order to trade off the conventional {\em spatial} sampling with {\em temporal} sampling. Also worth of note is the study in \cite{Antonsen}, which illustrates the transitioning from power-law to exponential time-decay of the power reflected by a SP-excited one-port cavity with ray-chaotic geometry.

Our interest in this subject was initially motivated by the speculation that pulse-reverberating (multi-echoing) ray-chaotic enclosures might provide spatio-temporal randomized field distributions of potential interest for wideband EM interference and/or EM compatibility testbeds. Some preliminary studies, first via an oversimplified ray analysis \cite{ICEAA99} and subsequently via rigorous numerical simulations \cite{APMag} based on the finite-difference-time-domain (FDTD) method \cite{Taflove}, confirmed the viability of this intuition. In this paper, we develop and validate a rigorous analytical framework based on the extension to the SP time-domain (TD) of the RPW model.

Accordingly, the rest of the paper is organized as follows. In Sec. \ref{Sec:Phys}, with specific reference to a Sinai-stadium-shaped enclosure, we outline the physical model and related ray-based and wavefield observables, highlighting the peculiarities of the late-time response. In Sec. \ref{Sec:Stat}, we develop a SP-RPW model which, in the parametric regime of interest, yields a spatio-temporal Gaussian random field whose statistics are amenable to analytical treatment.
In Sec. \ref{Sec:Num}, we validate the SP-RPW model against the spatio-temporal statistical observables estimated
from a full-wave numerical solution based on the FDTD method. Finally, in Sec. \ref{Sec:Concl}, we provide some brief concluding remarks and perspectives.

%
\begin{figure}
\begin{center}
\includegraphics [width=8.5cm]{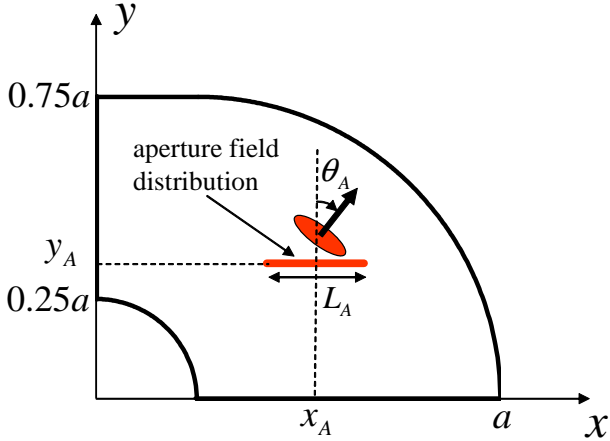}
\end{center}
\caption{Problem geometry. A 2-D (quarter) Sinai-stadium-shaped enclosure of characteristic size $a$ is excited by a SP wavepacket generated by a spatio-temporal aperture field distribution [cf. (\ref{eq:aperture})].}
\label{Figure1}
\end{figure}

\section{Physical Model}
\label{Sec:Phys}
\subsection{Geometry, Observables, and Notation}
The problem geometry is illustrated in Fig. \ref{Figure1}. We consider a two-dimensional (2-D) scenario featuring a closed, hollow enclosure with perfectly-electric-conducting (PEC) walls and $z-$invariant quarter-Sinai-stadium geometry \cite{Kudrolli} with characteristic size $a$. We assume the enclosure to be excited by a localized SP wavepacket generated by an (electric, $z-$directed) aperture field distribution along the $x-$axis, centered at $(x_A,y_A)$,
\beq
e_{Az}(x,t)=\Lambda(x-x_A)s\left[t-\frac{\sin\theta_A}{c_0}(x-x_A)\right],
\label{eq:aperture}
\eeq
with $\Lambda(x)$ denoting a spatial window localized within the interval $|x|<L_A/2$, $s(t)$ a SP waveform localized within the interval $|t|<T_s/2$, $c_0=1/\sqrt{\varepsilon_0\mu_0}$ the speed of light in vacuum, and $\theta_A$ a possible slant angle (see Fig. \ref{Figure1}). This yields a transverse-magnetic (TM) polarized EM field distribution
\beq
{\bf e}({\bf r},t)=e_z({\bf r},t)\hat{\bf z},~~{\bf h}({\bf r},t)=h_x({\bf r},t)\hat{\bf x}+h_y({\bf r},t)\hat{\bf y},
\eeq
with ${\bf r}\equiv x \hat{\bf x}+ y \hat{\bf y}$, and $\hat{\alphabf}$ denoting an $\alpha-$directed unit vector. Throughout the paper, boldface symbols identify vector quantities, lower- and upper-case symbols identify time-domain (TD) and frequency-domain (FD) observables, respectively, and a tilde (~$\tilde{}$~) identifies wavenumber-domain quantities, according to
\begin{subequations}
\begin{eqnarray}
f(t)&=&\frac{1}{2\pi}\int \limits_{-\infty}^{\infty} F(\omega) \exp\left(-i \omega t\right) d\omega,\\
F(\omega)&=&\int \limits_{-\infty}^{\infty} f(t) \exp\left(i \omega t\right) dt,
\end{eqnarray}
\label{eq:FT}
\end{subequations}
\begin{subequations}
\begin{eqnarray}
F({\bf r})&=&\frac{1}{4\pi^2}\iint \limits_{-\infty}^{~~~\infty}  
{\tilde F}({\bf k}) \exp\left(i {\bf k}\cdot {\bf r}\right) d{\bf k},\\
{\tilde F}({\bf k})&=&\iint \limits_{-\infty}^{~~~\infty}  
F({\bf r}) \exp\left(-i {\bf k}\cdot {\bf r}\right) d{\bf r}.
\end{eqnarray}
\end{subequations}

In view of the assumed lossless character, straightforward application of Poynting theorem \cite{Stratton} yields the conservation of the total EM energy (per unit length in $z$) injected in the enclosure through the aperture, given by 
\begin{eqnarray}
{\cal E}_{EM}&\equiv&
\iint\limits_{{\cal{D}}_E}  w_{EM}({\bf r},T_e) d{\bf r}\nonumber\\
&=&
\int\limits_{-\infty}^{T_e} dt 
\int\limits_{{\cal{D}}_A} 
{\bf e}_A(x,t)\times {\bf h}_A(x,t)\cdot \hat{\bf y} dx,
\label{eq:EEM}
\end{eqnarray}
where ${\cal{D}}_E$ and ${\cal{D}}_A$ indicate the enclosure and aperture domains, respectively, ${\bf e}_A, {\bf h}_A$ the EM field at the aperture,
\beq
w_{EM}({\bf r},t)
=
\frac{1}{2}
\left[
\varepsilon_0 {\bf e}({\bf r},t)\cdot {\bf e}({\bf r},t)+
\mu_0 {\bf h}({\bf r},t)\cdot {\bf h}({\bf r},t)
\right]
\label{eq:wEM}
\eeq
the EM energy density, and
\beq
T_e>\frac{T_s}{2}+\frac{L_A\sin{\theta_A}}{2c_0}
\label{eq:TT}
\eeq
guarantees the extinction of the aperture-field transient.

%
\begin{figure}
\begin{center}
\includegraphics [width=8.5cm]{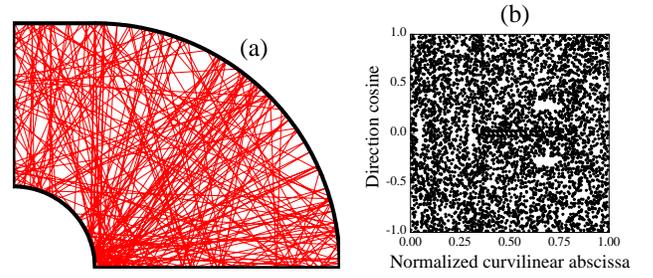}
\end{center}
\caption{(a) Typical space-filling ray trajectory (after 250 reflections) in a quarter-Sinai-stadium enclosure. (b) Pseudo phase-space map displaying, for 5000 reflections, the curvilinear abscissa [measured counterclockwise from the point of coordinates ($x=0, y=0.25a$), normalized to the total boundary length] vs. the direction cosine of the reflected ray.}
\label{Figure2}
\end{figure}

\subsection{Ray-Chaotic Behavior}
The quarter-Sinai-stadium geometry considered in our study, first introduced in \cite{Kudrolli}, consists of a quarter stadium with an off-centered quarter-circle indentation (Fig. \ref{Figure1}), and is basically a combination of the well-known Sinai (rectangle with central circle) \cite{Sinai} and Bunimovich (stadium) \cite{Bunimovich} billiard geometries. Similar to these canonical geometries, it exhibits ray-chaotic behavior but, unlike them, it is devoid of non-isolated periodic trajectories (``bouncing-ball'' modes). 

While a comprehensive study of the ray-chaotic dynamics is not in order here (see, e.g., \cite{APMag} for more details), Fig. \ref{Figure2} provides a basic illustration of the phenomenon. More specifically, Fig. \ref{Figure2}(a) displays the space-filling properties of a typical ray trajectory (after 250 reflections), while Fig. \ref{Figure2}(b) shows a pseudo phase-space diagram, namely, for each reflection, the curvilinear abscissa of the impact point vs. the direction cosine of the reflected ray (details in the caption). The practically uniform covering of the phase space confirms the visual impression from Fig. \ref{Figure2}(a) that (apart from zero-measure sets of launching conditions corresponding to isolated periodic trajectories) rays eventually approach almost every point arbitrarily closely and arbitrarily many times, with uniformly distributed arrival directions, when the number of reflections tends to infinity. This is markedly different from what observed in coordinate-separable geometries (e.g., rectangle, circle, ellipse), where ray trajectories tend to fill the enclosure in a {\em regular} fashion, with correspondingly sparsely populated phase spaces (see \cite{APMag} for more details).

Clearly, given their {\em purely kinematical} nature, the above considerations do not depend on the type (time-harmonic or SP) of excitation.

%
\begin{figure}
\begin{center}
\includegraphics [width=8.5cm]{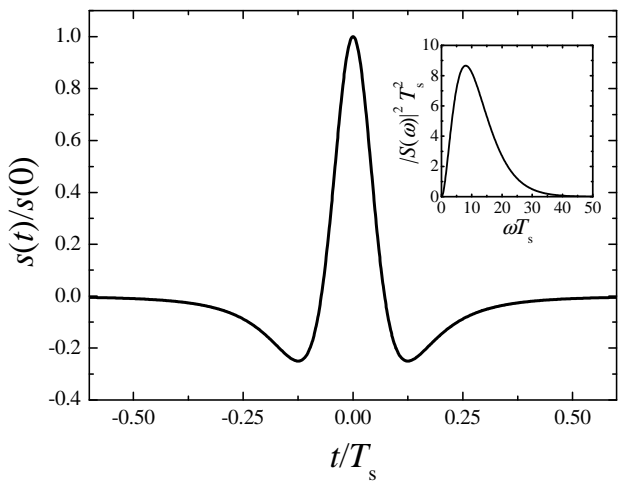}
\end{center}
\caption{SP (normalized) waveform considered in the FDTD simulations [cf. (\ref{eq:sdelta}) and (\ref{eq:Sdelta}), with $q=2$ and $\varsigma=0.25$]. The corresponding (normalized) energy spectral density is shown in the inset.}
\label{Figure3}
\end{figure}

\subsection{Full-Wave Modeling}
\label{Sec:FDTD}
For a preliminary qualitative understanding of the SP wavepacket propagation in the Sinai-stadium-shaped enclosure, as well as for numerical validation purposes (see Sec. \ref{Sec:Num}), we use the FDTD method \cite{Taflove}. In our simulations, the aperture field distribution in (\ref{eq:aperture}) is implemented as a ``hard source'' \cite{Taflove}, with a Gaussian spatial-window function
\beq
\Lambda(x)=\exp\left({-\frac{16 x^2}{L_A^2}}\right)
\label{eq:window}
\eeq
of effective width $L_A=a/6$, and the SP waveform shown in Fig. \ref{Figure3} (see Sec. \ref{Sec:workable} below for details) of effective duration $T_s=a/(20c_0)$, chosen so as to guarantee a significant spatio-temporal localization on the enclosure scale ($c_0T_s\ll a$).

The computational domain is discretized with spatial steps $\Delta_x=\Delta_y=a/1200$ [corresponding to $\sim 38$ points per center wavelength, and $\sim 15$ points per minimum wavelength, cf. the inset in Fig. \ref{Figure3}], and time step $\Delta_t=T_s/424\approx\Delta_x/(5\sqrt{2}c_0)$ (i.e., five-time below the Courant limit for numerical stability \cite{Taflove}).
The above spatio-temporal discretization parameters were chosen so as to guarantee a suitable tradeoff between computational affordability and accuracy (assessed via self-consistency tests based on energy conservation), as well as the need to provide a suitable number of spatio-temporal field samples for the statistical analysis in Sec. {\ref{Sec:Num}}.

\begin{figure*}
\begin{center}
\includegraphics [width=14cm]{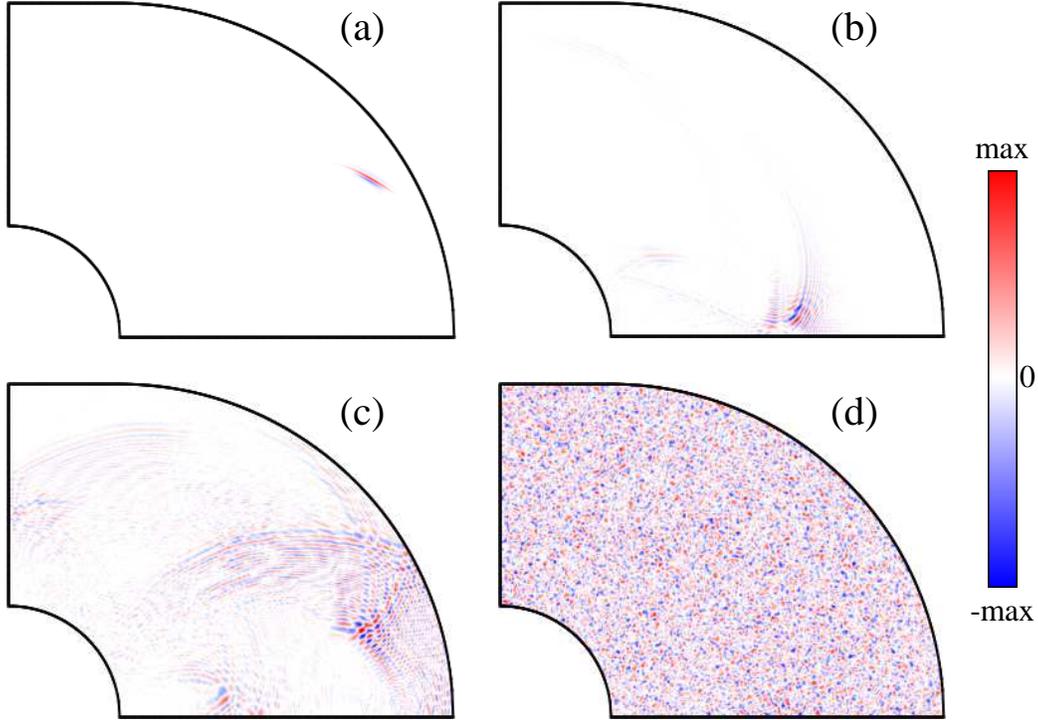}
\end{center}
\caption{FDTD-computed electric-field snapshots for the quarter-Sinai-stadium enclosure in Fig. \ref{Figure1} at (a) $t_0=3 T_s$, (b) $t_0=50 T_s$, (c) $t_0=100 T_s$, (d) $t_0=1000 T_s$. Hard-source spatio-temporal aperture field distribution as in (\ref{eq:aperture}), with $x_A=0.75a$, $y_A=0.25$, $\theta_A=30^o$; spatial window in (\ref{eq:window}) with $L_A=a/6$; SP waveform in Fig. \ref{Figure3} [cf. (\ref{eq:sdelta}) and (\ref{eq:Sdelta}), with $q=2$ and $\varsigma=0.25$] with duration $T_s=a/(20c_0)$.}
\label{Figure4}
\end{figure*}

Figure \ref{Figure4} shows a typical temporal evolution of the wavefield, via four instantaneous snapshots at representative times. While propagating inside the cavity and undergoing multiple reflections at the boundaries, the initial ``bullet-type'' wavepacket [Fig. \ref{Figure4}(a)] progressively loses its {\em shape} and spatio-temporal {\em localization}, and its energy is eventually spread across the enclosure in a uniform random-like fashion [Fig. \ref{Figure4}(d)]. 
In this regime, as already noted, {\em statistical modeling} may provide a better problem-matched description and parameterization, as it may effectively reproduce some {\em coarse} features (e.g., mean value and energy, correlation).
It should be noted, however, that such a description may fail satisfying the boundary conditions, and its use should be accordingly limited to  regions sufficiently far from the boundaries (e.g., in terms of the largest wavelength in the field spectrum).

\begin{figure}
\begin{center}
\includegraphics [width=8.5cm]{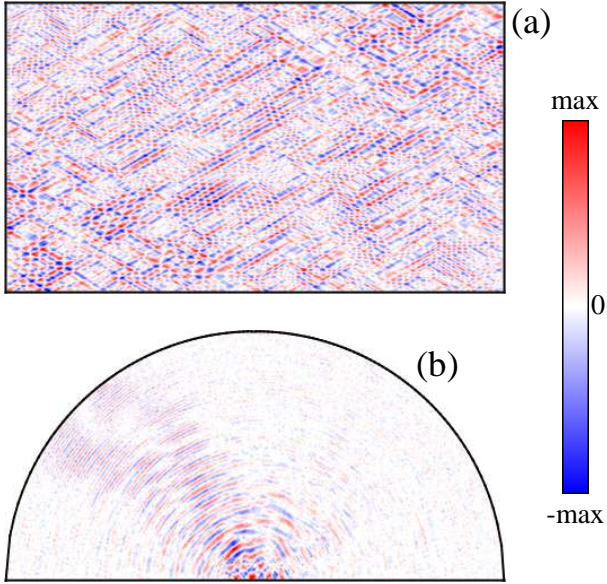}
\end{center}
\caption{As in Fig. \ref{Figure4}(d) (i.e., $t_0=1000 T_s$) but for geometry-separable ray-regular enclosures with (a) square and (b) half-circular shape, and same total area (not in scale).}
\label{Figure5}
\end{figure}

It is insightful to qualitatively compare the above behavior with what occurs in ``ray-regular'' (i.e., non-chaotic) coordinate-separable geometries. To this aim, Fig. \ref{Figure5} shows the late-time snapshots pertaining to a rectangular and a semi-circular enclosure of same are as the quarter-Sinai-stadium geometry. For the rectangular case [Fig. \ref{Figure5}(a)], a rather uniform and yet {\em regular} covering characterized by a checkerboard-type pattern is observed, which reflects the typical fixed-angle character of the associated ray trajectories \cite{APMag}. On the other hand, the field snapshot pertaining to the semi-circular geometry [Fig. \ref{Figure5}(b)] appears quite {\em disuniform}, reflecting the structure of typical annular-sector-filling ray trajectories bounded by circular caustics \cite{APMag}.

%

%

\section{Statistical Model}
\label{Sec:Stat}

\subsection{SP-RPW Superposition}
In \cite{Berry1}, Berry conjectured that the higher-order eigenstates in ray-chaotic quantum billiards should be statistically indistinguishable from a superposition of time-harmonic plane waves with random and uncorrelated directions of propagation, amplitudes, and phases. The RPW model, has been demonstrated (see, e.g., \cite{Gutz}--\cite{Stockmann}) to effectively capture the statistical properties of high-order wavefunctions of strongly chaotic billiards in the high-frequency regime (see, e.g., \cite{Berry2,Backer} for possible deviations due to bouncing-ball modes). In what follows, we extend the RPW model, perhaps for the first time, to the SP-TD scenario of interest here.

Let us consider a spatio-temporal domain of observation
\beq
r\equiv\left|{\bf r}\right|\le R,~~t_0\le t\le t_0+T,
\label{eq:domain}
\eeq
with $t_0$ sufficiently large so that the wavepacket energy has spread uniformly across the enclosure [as, e.g., in Fig. \ref{Figure4}(d)], and with $R\gg c_0T_s$ and $T\gg T_s$. Moreover, we make the assumption that the observation points {\bf r} in the domain are sufficiently far (i.e., several effective wavelengths or, equivalently, pulse widths $c_0T_s$ away) from the enclosure boundary (see the discussion in Sec. {\ref{Sec:FDTD}}).
We make the {\em ansatz} that the EM field distribution can be modeled in terms of a superposition of independent SP-RPWs of the form
\begin{subequations}
\begin{eqnarray}
e_z({\bf r},t)&\sim&
\sum_{n=1}^{N}a_n s\!\left(t-\kappabf_n \cdot {\bf r}-t_n\right),\\
{\bf h}({\bf r},t)&\sim&
\frac{1}{\mu_0}
\sum_{n=1}^{N}a_n s\!\left(t-\kappabf_n \cdot {\bf r}-t_n\right) \kappabf_n \times \hat{\bf z},
\end{eqnarray}
\label{eq:SP-RPW}
\end{subequations}
\!\!where $N$, $a_n$, $\kappabf_n$, and $t_n$ are independent random variables. More specifically, the SP-RPWs are assumed to follow a Poisson arrival process described by the probability function \cite{Middleton} 
\begin{subequations}
\beq
P_{T,R}\left(N\right)=\frac{\left({\bar N}_{T,R}\right)^N \exp\left(-N\right)}{N!},
\eeq
with ${\bar N}_{T,R}$ representing the average number of arrivals in the spatio-temporal domain of observation (\ref{eq:domain}). The amplitude coefficients $a_n$ are uncorrelated, zero-mean, unit-variance random variables,  
\beq
\left<a_n\right>=0,~~
\left<a_n a_m\right>= \delta_{nm},
\label{eq:an}
\eeq
with $\left<\cdot\right>$ denoting the statistical (ensemble) expectation value, $\delta_{nm}$ the Kronecker delta, and no specific assumption made on higher-order correlations. The ``normalized'' wavevectors are expressed by
\beq
\kappabf_n=c_0^{-1}\left(\cos\phi_n \hat{\bf x}+\sin\phi_n \hat{\bf y}\right),
\label{eq:kappan}
\eeq
where $\left\{\phi_n\right\}$ is a set of uniformly, independently distributed random variables within the interval $[-\pi,\pi]$. The time-delays $t_n$ are uniformly, independently distributed within the interval $[t_0-R/c_0,t_0+T+R/c_0]$, which guarantees causal connection with the spatio-temporal observation domain in (\ref{eq:domain}).
\end{subequations}

\subsection{Background Results on Poisson Impulse Noise Processes}
\label{Sec:Background}
In order to understand the statistical properties of the SP-RPW superpositions in (\ref{eq:SP-RPW}), we capitalize on a body of well-established technical results on Poisson impulse noise processes. In particular, following \cite{Middleton}, a critical nondimensional parameter for establishing the statistical structure is 
\beq
\gamma\equiv
\displaystyle{
\frac{{\bar N}_{T,R}T_s}{T+\displaystyle{\frac{2R}{c_0}}}},
\label{eq:gamma}
\eeq
which basically represents the average number of SP-RPWs per unit time at a fixed observation point.
Intuitively, small values of $\gamma$ (up to $\sim 10$ {\cite{Middleton}}) imply {\em weakly overlapping} SP-RPWs and, consequently, a strong dependence of the statistics on the SP waveform $s(t)$ as well as on the specific distributions of the random parameters $a_n$, $\kappabf_n$ and $t_n$, with a high probability associated with zero or small amplitude values \cite{Middleton}. This regime may somehow resemble the intermediate-time response of the enclosure, when the wavepacket energy is only partially spread across the enclosure [see, e.g., Fig. \ref{Figure4}(c)], and is not of direct interest for this investigation. In fact, the late-time response [cf. Fig. \ref{Figure4}(d)] of interest here is more intuitively associated with {\em strongly overlapping} SP-RPWs, and hence {\em large} values of $\gamma$. In this regime, as a consequence of the central limit theorem \cite{Middleton}, the statistics approach Gaussian behavior. More specifically {\cite{Middleton}}, nearly-Gaussian behaviors (with correction terms of order $\gamma^{-1/2}$ or $\gamma^{-1}$) should be expected for $\gamma$ values ranging from $\sim 10$ to $\sim 10^4$, whereas $\gamma \gtrsim 10^4$ should ensure Gaussian behavior 
 irrespective of the SP waveform shape and specific parameter distributions, with the parameter $\gamma$ appearing as a scale factor in the probability density function (PDF). In this regime, the (insofar unknown) parameter ${\bar N}_{T,R}$ may be related to the actual physical scenario via simple energy considerations (see Sec. {\ref{Sec:Link}} below).

\subsection{Spatio-Temporal Gaussian Random Field}
It follows from the considerations above that the late-time SP response of our ray-chaotic enclosure may be effectively modeled by a spatio-temporal Gaussian random field \cite{Le}. Accordingly, given a $K$-dimensional array of spatio-temporal field samples
\beq
\underline{e_z}=\left[e_z({\bf r}_1,{\cal T}_1)~~e_z({\bf r}_2,{\cal T}_2)~~...~~e_z({\bf r}_K,{\cal T}_K) \right]^T,
\eeq
with $^T$ denoting the transpose, the associated joint PDF is given by \cite{Le}
\begin{subequations}
\beq
p(\underline{e_z})=\frac{1}{(2\pi)^K\det({\underline {\underline \Sigma}})}\exp
\left[
-\frac{1}{2}\left(\underline{e_z}-{\underline \mu}\right)^T\cdot {\underline {\underline \Sigma}}^{-1}\cdot 
\left(\underline{e_z}-{\underline \mu}\right)
\right],
\label{eq:PDF}
\eeq
with 
\beq
{\underline \mu}=\left<\underline{e_z}
\right>,~~{\underline {\underline \Sigma}}=\left<
\left(
\underline{e_z}-{\underline \mu}
\right)\cdot
\left(
\underline{e_z}-{\underline \mu}
\right)^T
\right>
\eeq
\end{subequations}
representing the expectation-value vector and covariance matrix, respectively. It can readily be shown (see Appendix A for details) that
\beq
\left<e_z({\bf r},t)\right>=0,
\label{eq:mean}
\eeq
and therefore the Gaussian joint PDF in (\ref{eq:PDF}) is completely determined by the knowledge of the spatio-temporal correlation 
\beq
c(\rhobf,\tau)=\left<e_z\left({\bf r},t\right) e_z\left({\bf r}+\rhobf,t+\tau\right)\right>,
\label{eq:corr}
\eeq
where stationarity in space and time (i.e., independence on ${\bf r}$ and $t$) is anticipated. Alternatively, the correlation in (\ref{eq:corr}) may also be calculated via a spectral (Wiener-Khinchin-type) representation \cite{Le},
\beq
c(\rhobf,\tau)=\frac{1}{8\pi^3}\int\limits_{-\infty}^{\infty} d\omega 
\iint\limits_{-\infty}^{~~~\infty} d{\bf k} 
~{\tilde C}_e({\bf k},\omega)
\exp\left[
i\left(
{\bf k}\cdot\rhobf-i\omega \tau
\right)
\right],
\label{eq:ce}
\eeq
where ${\tilde C}_e({\bf k},\omega)$ is the power spectral density (PSD). Simple heuristic arguments would suggest a generalization of the known FD results \cite{Berry1} in terms of a separable PSD, 
\beq
{\tilde C}({\bf k},\omega)=\Xi \left|
S(\omega)
\right|^2
\delta\left({\bf k}\cdot{\bf k}-\frac{\omega^2}{c_0^2}
\right),
\label{eq:PSD}
\eeq
where $\Xi$ is a multiplicative constant, the Dirac-delta spectral kernel enforces the plane-wave dispersion relation in vacuum, and $|S(\omega)|^2$ accounts for the energy spectral density of the SP waveform $s(t)$. In view of the assumed structure, the wavevector integral in (\ref{eq:ce}) can be performed analytically via a simple change of variables to cylindrical coordinates,
${\bf k}=\omega c_0^{-1}(\cos\varphi\hat{\bf x}+\sin\varphi\hat{\bf y})$, $\rhobf=\rho(\cos\varphi_0\hat{\bf x} +\sin\varphi_0\hat{\bf y})$, yielding
\begin{eqnarray}
\!\!\!\!c(\rho,\tau)\!\!\!\!\!&=&\!\!\!\!\!\frac{\Xi}{8\pi^3}
\int\limits_{-\infty}^{\infty} d\omega \left|
S(\omega)
\right|^2 \exp\left(-i\omega\tau\right)\nonumber\\
&\times&\int\limits_{-\pi}^{\pi} 
\exp\left[i \frac{\omega\rho}{c_0}\cos\left(\varphi-\varphi_0\right)
\right] d\varphi\nonumber\\
&=&
\frac{\Xi}{4\pi^2}
\int\limits_{-\infty}^{\infty} |S(\omega)|^2
J_0\!\left(\frac{\omega \rho}{c_0}\right)
\exp\left(-i \omega \tau\right)
d\omega,
\label{eq:ce2}
\end{eqnarray}
where $J_0$ denotes a zeroth-order Bessel function \cite{Watson}. We note that the result in (\ref{eq:ce2}) does not depend on the {\em direction} of the spatial displacement vector ${\rhobf}$, thereby implying {\em isotropy} of the spatio-temporal Gaussian random field. Moreover, Eq. (\ref{eq:ce2}) reproduces the well-known $J_0$-type FD form \cite{Berry1} in the time-harmonic limit.

As shown in Appendix B, the above heuristic-based result is consistent with first-principle calculations [i.e., with (\ref{eq:corr}) and (\ref{eq:SP-RPW})]. In particular, as also shown in Appendix B, we can relate the multiplicative constant $\Xi$ in (\ref{eq:ce2}) to the critical parameters in the SP-RWP model, viz.,
\beq
\Xi=\frac{\pi\gamma}{T_s},
\label{eq:FP}
\eeq
whereby the anticipated appearance of the critical parameter $\gamma$ in (\ref{eq:gamma}) as a scaling factor in the Gaussian PDF. 

\subsection{Link with Underlying Physics}
\label{Sec:Link}
A link between the SP-RPW model and the underlying physics can be established via simple energy considerations. First, via straightforward vector algebra, we note that, in spite of the random character of the normalized wavevector ${\kappabf}_n$ in (\ref{eq:SP-RPW}), the quantity
\beq
\left(\kappabf_n \times \hat{\bf z}\right)\cdot\left(\kappabf_n \times \hat{\bf z}\right)=c_0^{-2}
\eeq
is {\em always deterministic}. This implies that the spatio-temporal correlation for the magnetic field is simply related to that of the electric field in (\ref{eq:corr}) via
\beq
\left<{\bf h}\left({\bf r},t\right)\cdot {\bf h}\left({\bf r}+\rhobf,t+\tau\right)\right>= \frac{c(\rho,\tau)}{\eta_0^2},
\label{eq:ch}
\eeq
where $\eta_0=\sqrt{\mu_0/\varepsilon_0}$ is the vacuum characteristic impedance. Moreover, Eq. (\ref{eq:ch}) also establishes a simple connection with the expectation value of the EM energy density in (\ref{eq:wEM}),
\beq
\left<w_{EM}({\bf r},t)\right>=\varepsilon_0 c(0,0),
\eeq
which, in view of spatio-temporal stationarity, is in turn simply related to the total EM energy (per unit length) in (\ref{eq:EEM}) via
\beq
\left<w_{EM}({\bf r},t)\right> =\frac{{\cal E}_{EM}}{{\cal A}_E},
\eeq
where ${\cal A}_E$ is the area of the enclosure. Accordingly, we can rewrite the spatio-temporal correlation in (\ref{eq:ce2}) as
\begin{eqnarray}
c(\rho,\tau)= 
\left(\frac{{\cal E}_{EM}}{\varepsilon_0 {\cal A}_E}\right)
\int\limits_{-\infty}^\infty  
|{\bar S}(\omega)|^2
J_0\!\left(\frac{\omega \rho}{c_0}\right)
\exp\left(-i \omega \tau\right)
d\omega,
\label{eq:ce1}
\end{eqnarray}
where
\beq
|{\bar S}(\omega)|^2=\frac{|S(\omega)|^2}{\int\limits_{-\infty}^\infty  
|S(\omega)|^2
d\omega}
\eeq
represents the normalized energy spectral density of the SP waveform. The expression in (\ref{eq:ce1}) relates the statistical properties of the SP-RPW model to simple geometrical and energy invariants of the physical model.

\subsection{Analytically Workable Examples}
\label{Sec:workable}
It is instructive to work out a class of examples for which the integral in (\ref{eq:ce1}) can be calculated analytically in closed form. To this aim, we consider a class of SP waveforms
\beq
s(t)=\mbox{Re}\left[\stackrel{+}{\delta}~\!\!\!^{(q)}\!\left(t-i \frac{\varsigma T_s}{2}\right)\right],
\label{eq:sdelta}
\eeq
where $\varsigma$ is a scaling parameter (chosen so as to ensure that the effective waveform duration is $T_s$ within a suitably small error), the superfix $^{(q)}$ indicates the $q-$th derivative, and 
\begin{eqnarray}
\stackrel{+}{\delta}\!(t)&=&\frac{1}{\pi}\int\limits_{0}^{\infty}\exp\left(i\omega t\right)d\omega\nonumber\\
&=&
\left\{
\begin{array}{ll}
\delta(t)+{\cal P}\!\left(\displaystyle{\frac{1}{i\pi t}}\right),~~~\mbox{Im}(t)=0\\
\displaystyle{\frac{1}{i\pi t}},~~~\mbox{Im}(t)<0
\end{array}
\right.
\end{eqnarray}
is the analytic delta function \cite{Heyman} (with ${\cal P}$ denoting the principal value). It can readily be shown that
\beq
S(\omega)=(-i\omega)^q\exp\left(-|\omega| \frac{\varsigma T_s}{2}\right),
\label{eq:Sdelta}
\eeq
which, substituted in (\ref{eq:ce1}), yields (cf. Eq. (2.12.8.4) in \cite{Prudnikov})
\begin{eqnarray}
c(\rho,\tau)&=&
\frac{(\varsigma T_s)^{2q+1}{\cal E}_{EM}}{\varepsilon_0 {\cal A}_E}
\mbox{Re}
\left\{
\left[
\frac{\rho}{c_0\xi}
\left(1+\xi^2\right)^{\frac{1}{2}}
\right]
^{-(2q+1)}
\right.\nonumber\\
&\times&\Biggl.
P_{2q}\left[
\left(1+\xi^2
\right)^{-\frac{1}{2}}
\right]
\Biggr\},
\label{eq:cex}
\end{eqnarray}
where $P_m$ denotes the $m$th-degree Legendre polynomial (cf. Sec. 8 in \cite{Abramowitz}), and
\beq
\xi=\frac{\rho}{c_0\left(\varsigma T_s+i\tau\right)}
\eeq
is a complex, nondimensional spatio-temporal variable. Particularly simple and insightful is the special case $q=0$ (analytic delta), for which
\begin{subequations}
\beq
c(\rho,\tau)=
\left(\frac{\varsigma T_s{\cal E}_{EM}}{\varepsilon_0 {\cal A}_E}\right)
\mbox{Re}
\left\{
\left[
\frac{\rho^2}{c_0^2}+
\left(
\varsigma T_s+i\tau
\right)^2
\right]^{-1/2}
\right\},
\eeq
whose 1-D cuts at $\tau=0$ and $\rho=0$ can be written as
\beq
c(\rho,0)=
\frac{\rho_c{\cal E}_{EM}}{\varepsilon_0 {\cal A}_E \sqrt{3\rho^2+\rho_c^2}},
\eeq
\beq
c(0,\tau)=
\frac{T_c^2 {\cal E}_{EM}}{\varepsilon_0 {\cal A}_E \left(\tau^2+T_c^2\right)},
\eeq
\label{eq:simplest}
\end{subequations}
\!\!with $\rho_c=\sqrt{3} c_0 \varsigma T_s$ and $T_c=\varsigma T_s$ representing the correlation length and time (3dB below), respectively. This simple example completes the illustration of how the statistical properties of the late-time field distribution, modeled as a spatio-temporal Gaussian random field, are related to (and may be potentially controlled via) a {\em minimal} number of generic geometrical and physical parameters.

\section{Numerical Validation}
\label{Sec:Num}

\subsection{Generalities and Observables}
We now move on to validating the SP-RPW model developed in Sec. \ref{Sec:Stat} against the statistical observables that can be {\em directly} estimated from the FDTD-computed late-time wavefield distributions [cf. Fig. \ref{Figure4}(d)]. In order to facilitate direct comparison with the theoretical predictions, we introduce a suitably normalized field,
\beq
{\bar e}_z({\bf r},t)=\sqrt{\frac{\varepsilon_0 {\cal A}_E}{{\cal E}_{EM}}} ~e_z({\bf r},t),
\label{eq:normalized}
\eeq
for which the predicted statistics are {\em standard} (i.e., zero-mean, unit-variance) Gaussians. In (\ref{eq:normalized}), ${\cal A}_E=(3+8\pi)a^2/16$ (cf. Fig. \ref{Figure1}), the wavelfield $e_z$ is computed as detailed in Sec. \ref{Sec:FDTD}, and the EM energy (per unit length) ${\cal E}_{EM}$ is computed numerically via spatial integration of the EM energy density [cf. (\ref{eq:EEM})--(\ref{eq:TT})]. Note that, in order to avoid possible direct-current field offset issues in the FDTD simulations \cite{Taflove}, we prefer to work with the zero-mean SP waveform in Fig. \ref{Figure3}, rather than the simplest analytic-delta ($q=0$) case in (\ref{eq:simplest}). 

Assuming ergodicity, the estimation of the statistical (ensemble) observables of interest is based on {\em single} spatial or temporal wavefield realizations. Moreover, in order to reveal possible variations (in time, and across the enclosure), we are interested in {\em local} statistics, and accordingly consider observation domains with sufficiently {\em small} size (compared with the overall spatio-temporal observation domain), and yet sufficiently {\em large} (on the SP wavepacket scale) so as to guarantee meaningful statistics. More specifically, unless otherwise stated, local temporal statistics are computed over time-series of size $20 T_s$ (i.e., 8478 samples) at a fixed observation point, while local spatial statistics are computed over square domains of size $4c_0T_s\times 4c_0T_s$ (i.e., 57121 samples) around a given point.

%
\begin{figure}
\begin{center}
\includegraphics [width=8.5cm]{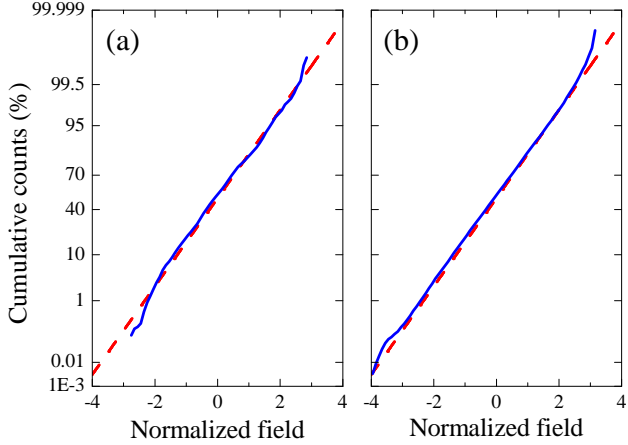}
\end{center}
\caption{Gaussian probability plots (blue-solid curves) for local (a) temporal and (b) spatial statistics of the normalized field in (\ref{eq:normalized}) estimated from a single FDTD-computed realization. Local temporal statistics are estimated observing the wavefield at a fixed point ($x_c=0.602a, y_c=0.121a$) over an interval of duration $20 T_s$ (i.e., $M=8478$ samples) starting at $t_0=1000 T_s$ (estimated mean and variance are $-0.0035$ and $1.07$, respectively). Local spatial statistics are estimated over a $4c_0T_s\times 4c_0T_s$ square domain (i.e., $M=57121$ samples) centered at point ($x_c,y_c$) at time $t_0=1000 T_s$ (estimated mean and variance are $-0.0236$ and $1.057$, respectively). As a reference, also shown (red-dashed straight line) is the behavior of a standard (zero-mean, unit-variance) Gaussian distribution.}
\label{Figure6}
\end{figure}

%
\begin{figure}
\begin{center}
\includegraphics [width=8.5cm]{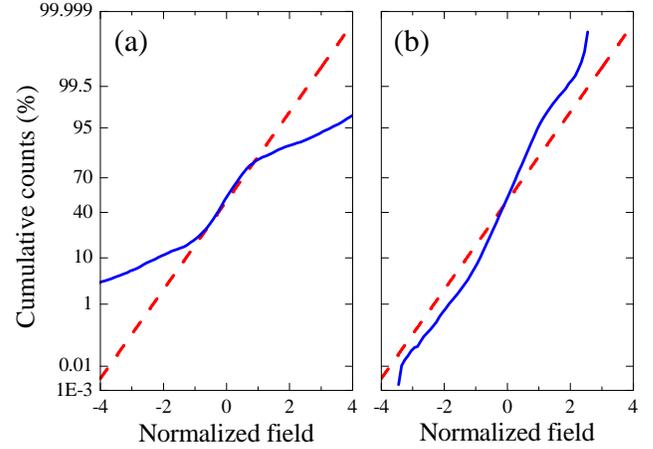}
\end{center}
\caption{As in Fig. \ref{Figure6}, but for earlier-time ($t_0=100 T_s$) observations. (a) Local temporal statistics (estimated mean and variance are $0.0018$ and $3.389$, respectively). (b) Local spatial statistics (estimated mean and variance are $-0.00649$ and $0.448$, respectively).}
\label{Figure7}
\end{figure}

%
\begin{figure}
\begin{center}
\includegraphics [width=8.5cm]{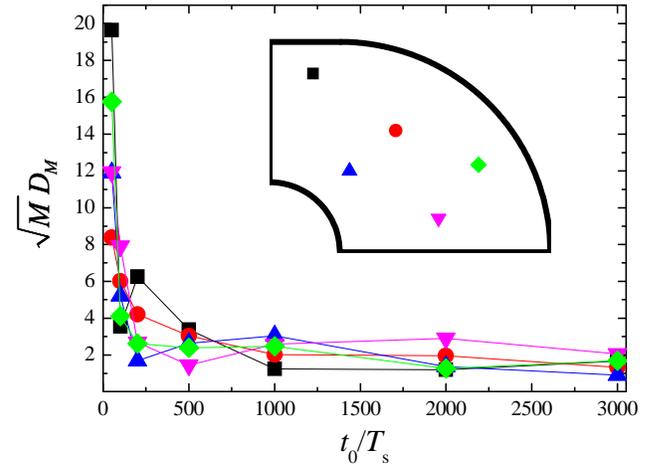}
\end{center}
\caption{Normalized KS distance in (\ref{eq:KS1}) pertaining to temporal field observations at five randomly-chosen points (shown as different markers in the insets) over an interval of duration $20 T_s$ (i.e., $M=8478$ samples) for different starting times $t_0$.}
\label{Figure8}
\end{figure}

\subsection{Representative Results}
For a first qualitative analysis, we show in Fig. \ref{Figure6} the Gaussian probability plots pertaining to point ($x_c=0.602a, y_c=0.121a$) (magenta down-triangle marker in the inset of Fig. \ref{Figure8}) at a {\em late time} $t_0=1000T_s$, when the SP wavepacket energy has spread uniformly across the enclosure [cf. Fig. \ref{Figure4}(d)]. More specifically, with reference to the local temporal [Fig. \ref{Figure6}(a)] and spatial [Fig. \ref{Figure6}(b)] distributions, the plots display the cumulative counts pertaining to the normalized field in (\ref{eq:normalized}) using a Gaussian probability scale, so that standard-Gaussian-distributed data sets would yield a reference straight line (shown as red-dashed). This facilitates direct visual assessment of the Gaussianity of the data, or otherwise the presence of short/long tails and/or left/right skewness.
As evident from Fig. \ref{Figure6}, at the chosen late time, both the temporal and spatial wavefield distributions 
follow quite closely the reference standard-Gaussian behavior predicted by the SP-RPW model.

%
\begin{figure}
\begin{center}
\includegraphics [width=8.5cm]{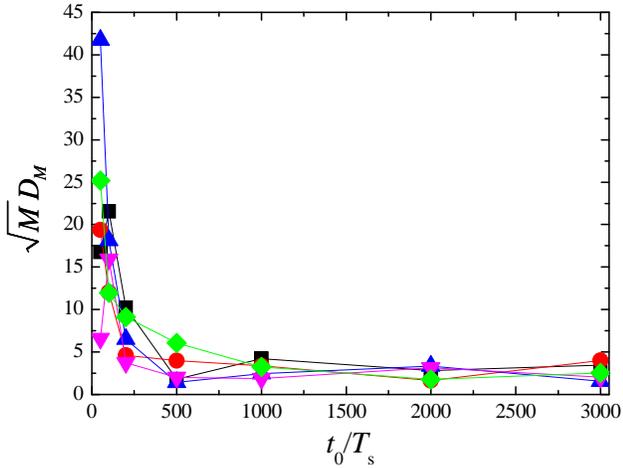}
\end{center}
\caption{As in Fig. \ref{Figure8}, but for spatial observations over $4c_0T_s\times 4c_0T_s$ square domains (i.e., $M=57121$ samples) centered at the same randomly-chosen points.}
\label{Figure9}
\end{figure}

Figure \ref{Figure7} shows the Gaussian probability plots pertaining to an {\em earlier-time} [$t=100T_s$, cf. Fig. \ref{Figure4}(c)] response at the same observation point, which turn out to be {\em markedly non-Gaussian} (see also the estimated mean and variance values given in the caption). Diverse types of distributions, not shown here for brevity, were observed for different observation points, confirming the general trend of markedly non-Gaussian behavior with significant parameter dependence on the observation point.

For a more quantitative assessment and compact visualization of the temporal and spatial fluctuations of the local statistics, it is expedient to introduce a measure of distribution distance, namely, the Kolmogorov-Smirnov (KS) \cite{KS},
\beq
D_M = \sup_X \left|\Phi_M(X) - \Phi(X) \right|,
\label{eq:KS1}
\eeq
where $\Phi_M$ is the cumulative density function (CDF) numerically estimated over $M$ samples, and
\beq
\Phi(X)=\frac{1}{2}\left[
1+\mbox{erf}\left(\frac{X}{\sqrt{2}}\right)
\right]
\eeq
is the standard Gaussian CDF.
Figure \ref{Figure8} shows the evolution of the (normalized) KS distance in (\ref{eq:KS1}) pertaining to the local temporal statistics at five randomly-chosen observation points (shown as different markers in the inset) with distances $\gtrsim 2 c_0T_s$ from the enclosure boundary, as a function of the observation time.
It is evident that, starting from rather large and strongly position-dependent values at early times, the KS distance progressively decreases with increasing observation times, converging to much smaller (about an order of magnitude) and uniform values for sufficiently late times. For quantitative reference, we recall that, in typical KS-based goodness-of-fit tests, the null hypothesis (i.e., Gaussian distribution) is rejected with a 0.01 significance level (i.e., 99\% confidence level) if \cite{KS}
\beq
\sqrt{M}D_M\gtrsim 1.63.
\eeq
Therefore, although a strict goodness-of-fit assessment is not in order here, we can safely conclude that the SP-RPW-based Gaussian prediction  seems to model reasonably well, and fairly uniformly across the enclosure, the late-time local temporal statistics. Similar observations and conclusions hold for the local spatial statistics over neighborhoods of the same observation points, as shown in Fig. \ref{Figure9}.

%
\begin{figure*}
\begin{center}
\includegraphics [width=14cm]{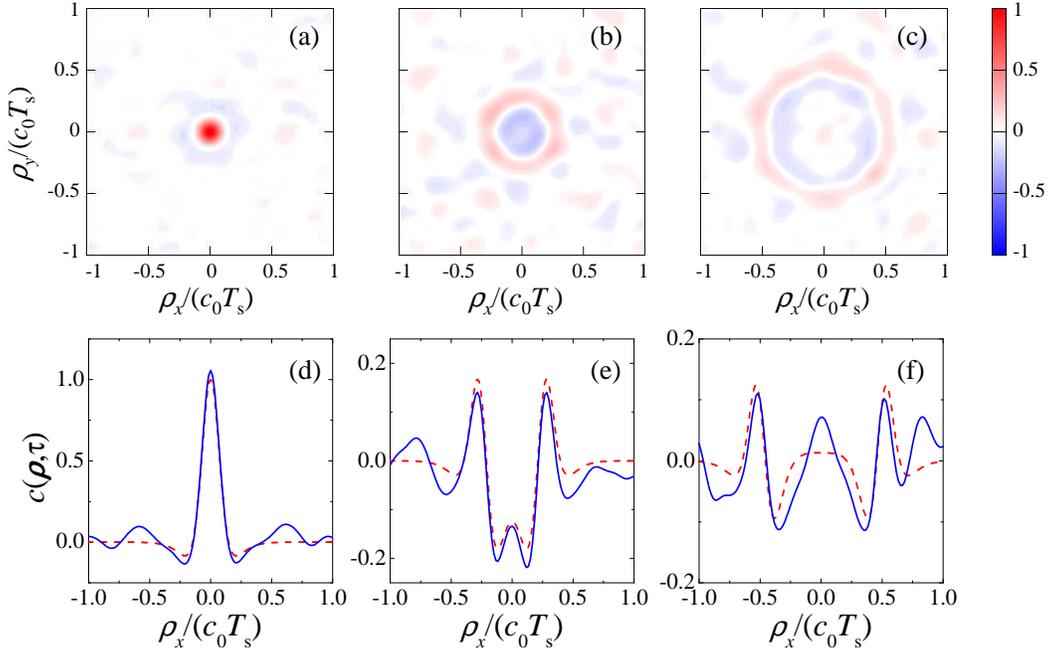}
\end{center}
\caption{Top panels: (a), (b), (c) Spatio-temporal correlation snapshots (as a function of $\rhobf=\rho_x \hat{\bf x}+\rho_y \hat{\bf y}$) at time $t_0=1000T_s$, estimated over a $4c_0T_s\times 4c_0T_s$ square domain (i.e., $M=57121$ samples) centered at point ($x_c=0.602a, y_c=0.121a$), for $\tau=0, 0.25T_s, 0.5T_s$, respectively. Bottom panels: (d), (e), (f) Corresponding 1-D cuts (blue-solid) at $\rho_y=0$ compared with SP-RPW prediction (red-dashed) from (\ref{eq:cex}) (with $q=2$ and $\varsigma=0.25$).}
\label{Figure10}
\end{figure*}

\begin{figure}
\begin{center}
\includegraphics [width=8.5cm]{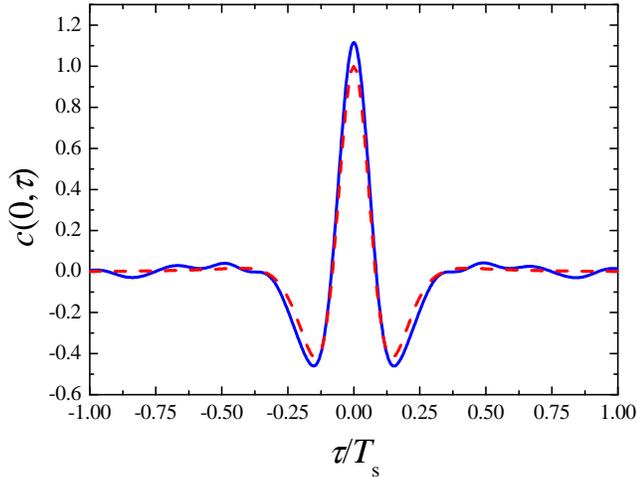}
\end{center}
\caption{Temporal ($\rhobf=0$) correlation (blue-solid) estimated from a wavefield observation at 
a fixed point ($x_c=0.602a, y_c=0.121a$) over an interval of duration $200 T_s$ (i.e., $M=84780$ samples) starting at $t_0=1000 T_s$, compared with SP-RPW prediction (red-dashed) from (\ref{eq:cex}) (with $q=2$ and $\varsigma=0.25$).}
\label{Figure11}
\end{figure}

Figures \ref{Figure10}(a)--\ref{Figure10}(c) show three representative snapshots of the late-time spatio-temporal correlation [estimated within a neighborhood of point ($x_c=0.602a, y_c=0.121a$)] for different values of the time lag $\tau$, with the corresponding 1-D cuts compared in Figs. \ref{Figure10}(d)--\ref{Figure10}(f) with the theoretical predictions from the SP-RPW model [cf. (\ref{eq:cex}) with $q=2$ and $\varsigma=0.25$].
Apart from weak fluctuations attributable to edge effects induced by the limited extent ($4c_0T_s\times4c_0T_s$) of the observation domain, the dominant features appear rather {\em isotropic}, and well captured by the SP-RPW prediction.
Finally, Fig. \ref{Figure11} shows the late-time temporal (i.e., zero spatial displacement) correlation estimated at a fixed observation point over an interval of duration $200 T_s$, compared with the SP-RPW prediction. Once again, a quite good agreement is observed. Qualitatively similar results were consistently observed for different observation points across the enclosure.

The above results clearly demonstrate that the SP-RPW model developed in Sec. \ref{Sec:Stat} effectively captures and parameterizes the essential statistical features of the irregular random-like regime that characterizes the late-time response of our ray-chaotic enclosure in the SP regime.

It is interesting to compare our SP-TD results with the well-known time-harmonic case. In that case, the local spatial statistics (estimated over regions of size much larger than the wavelength, but much smaller than the enclosure size) of high-order eigenfields approach Gaussian behavior, with spatial correlation approaching a Bessel function ($J_0$) form, which is well captured by time-harmonic RPW superpositions {\cite{Berry1}}. In our SP-TD case, the late-time local spatio-temporal statistics (estimated over regions of size much larger than the SP scale, but much smaller than the observation domain) approach Gaussian behavior, with spatio-temporal correlation depending on the SP waveform and coarse (geometrical and energy) parameters via ({\ref{eq:ce1}}).

\section{Conclusions and Perspectives}
\label{Sec:Concl}
In this paper, we have developed and validated a SP-RPW theoretical framework for the statistical modeling of the SP wavepacket propagation in a ray-chaotic enclosure. 
With specific reference to the quarter-Sinai-stadium geometry, our results confirm that the late-time response of the enclosure can be effectively modeled as a spatio-temporal random Gaussian field, whose parameters are simply related to coarse geometrical quantities, energy invariants, and the SP waveform.

The peculiar (and, to a certain extent, controllable) features of such wavefield distribution may find important applications to  wideband EM interference and/or EM compatibility testbeds (whereby an equipment under test may be subjected to a ``pulse shower'' illumination with characteristics largely independent of its location, orientation, shape and constitutive properties), as well as the synthetic emulation of multi-path wireless channels.
Within this framework, our results, based on the idealized assumption of the absence of losses, may be extended to a more realistic low-loss scenario by assuming {\em adiabatic} variations of the statistics as a consequence of the {\em slow} (on the SP time-scale) energy decay. Also of great interest is the exploration of electronic-stirring mechanisms based on narrowband-excited reverberating chambers with ray-chaotic geometries.

%

\section*{Acknowledgment}
This paper is dedicated to the dear memory of Professor Leo Felsen, {\em magister et amicus}, who first posed the problem
addressed here, and actively participated in our early
investigations \cite{APMag}.


\section*{Appendix A\protect\\
Pertaining to Eq. (\ref{eq:mean})}
In order to calculate the electric-field expectation value
\beq
\left<e_z\left({\bf r},t\right)\right>\sim
\left<
\sum_{n=1}^{N}a_n s\!\left(t-\kappabf_n \cdot {\bf r}-t_n\right)
\right>,
\label{eq:TDmean}
\eeq
we recall Wald's identity \cite{Wald} on the expectation value of the sum of a random number $N$ of finite-mean, identically distributed random variables $X_n$ (independent of $N$), viz.,
\beq
\left<
\sum_{n=1}
^{N}X_n\right>
=\left<N\right> \left<X_n\right>.
\label{eq:Wald}
\eeq
Straightforward application of (\ref{eq:Wald}) to (\ref{eq:TDmean}), and simple factorization (in view of the assumed statistical independence), yields
\beq
\left<e_z\left({\bf r},t\right)\right>\sim
{\bar N}_{T,R}\left<a_n\right>
\left<
s\!\left(t-\kappabf_n \cdot {\bf r}-t_n\right)\right>,
\eeq
from which Eq. (\ref{eq:mean}) follows immediately recalling (\ref{eq:an}).

\section*{Appendix B\protect\\ 
Pertaining to Eq. (\ref{eq:FP})}
The first-principle calculation of the spatio-temporal correlation via (\ref{eq:corr}) with (\ref{eq:SP-RPW}) is most easily carried out in the FD, using the spectral representation [via (\ref{eq:FT})] of the SP-RPW model
\beq
e_z\left({\bf r},t\right)\sim
\frac{1}{2\pi}
\sum_{n=1}^{N}a_n
\int\limits_{-\infty}^{\infty} S(\omega)
\exp\left[i \omega
\left(
\kappabf_n
\cdot {\bf r}
+t_n-t
\right)
\right]d\omega.
\label{eq:eFD}
\eeq
It is also expedient to rewrite (\ref{eq:corr}) as
\beq
c(\rhobf,\tau)=\left<e_z\left({\bf r},t\right) 
e_z^*\left({\bf r}+\rhobf,t+\tau\right)\right>,
\label{eq:corrc}
\eeq
where the ``dummy'' (in view of the real-valued wavefield) complex conjugation simplifies the following technical developments. By substituting (\ref{eq:eFD}) in (\ref{eq:corrc}), and rearranging terms, we obtain
\begin{eqnarray}
c(\rhobf,\tau)
&\sim&\frac{1}{4\pi^2}\left<
\sum_{n,m=1}^{N} a_n a_m
\iint\limits_{-\infty}^{~~~\infty} d\omega_1 
d\omega_2
 S(\omega_1)S^*(\omega_2)
 \right.
\nonumber\\ 
&\times&
\exp
\left[
i \omega_1 
\left(
\kappabf_n \cdot {\bf r}
+t_n
\right)\right]\nonumber\\
&\times&
\exp\left[-i \omega_2 
\left(
\kappabf_m\cdot {\bf r}
+\kappabf_m\cdot \rhobf
+t_m
\right)
\right]
\nonumber\\
&\times&
\Biggl.
\exp\left[
-i \left(\omega_1-\omega_2\right)t
+i\omega_2 \tau
\right]
\Biggr>.
\label{eq:cc3}
\end{eqnarray}
Factorizing out (in view of the assumed statistical independence) the correlation term $<a_n a_m>$ and recalling (\ref{eq:an}), Eq. (\ref{eq:cc3}) yields 
\begin{eqnarray}
c(\rhobf,\tau)&\sim&\frac{1}{4\pi^2}\left<
\sum_{n=1}^{N}
\iint\limits_{-\infty}^{~~~\infty} d\omega_1 
d\omega_2
 S(\omega_1)S^*(\omega_2)
 \right.
 \nonumber\\
&\times&
 \exp\left[
 i\left(\omega_1-\omega_2
 \right)t_n
 \right]
\nonumber\\ 
&\times&
\exp
\left[
-i\left(\omega_2-\omega_1\right)\left(t-\kappabf_n\cdot{\bf r}\right)
\right]\nonumber\\
&\times&
\Biggl.
\exp\left[i\omega_2
\left(\tau-\kappabf_n\cdot\rhobf
\right)
\right]
\Biggr>.
\end{eqnarray}
Applying once again Wald's identity (\ref{eq:Wald}), and recalling that
\begin{eqnarray}
&&\hspace*{-2cm}
\left<
\exp
\left[
i\left(
\omega_1 - \omega_2
\right)t_n\right]\right>\nonumber\\
&=&
\frac{1}{\left(T+\frac{2R}{c_0}\right)}\nonumber\\
&\times&
\int\limits_{t_0-\frac{R}{c_0}}^{t_0+T+\frac{R}{c_0}}
\exp
\left[
i\left(
\omega_1-\omega_2
\right)t_n\right] dt_n\nonumber\\
&=&
\frac{\exp
\left[
i\left(
\omega_1-\omega_2
\right)t_0\right]}{\left(T+\frac{2R}{c_0}\right)}\nonumber\\
&\times&
\int\limits_{-\frac{R}{c_0}}^{T+\frac{R}{c_0}}
\exp
\left[
i\left(
\omega_1-\omega_2
\right)t_n\right] dt_n\nonumber\\
&\sim&
\frac{\pi\exp
\left[
i\left(
\omega_1-\omega_2
\right)t_0\right]}{\left(T+\frac{2R}{c_0}\right)} ~\delta\left(\omega_1-\omega_2\right),
\end{eqnarray}
with the last asymptotic equality valid for $R\gg c_0T_s$ and $T\gg T_s$ (so as to approximately extend the integration limits to $\mp \infty$), we obtain
\begin{eqnarray}
\!\!\!\!\!\!c(\rhobf,\tau)\!\!\!\!&\sim&\!\!\!\!
\frac{{\bar N}_{T,R}}{4\pi\left(T+\frac{2R}{c_0}\right)}
\nonumber\\
\!\!\!\!\!\!&\times&\!\!\!\!\!\!
\int\limits_{-\infty}^{\infty} \left|S(\omega)\right|^2
\left<
\exp
\left(i\omega
\kappabf_n\cdot\rhobf
\right)
\right>\exp\left(-i\omega\tau\right)
d\omega.
\end{eqnarray}
The expectation value with respect to $\kappabf_n$ can be easily computed
by letting $\rhobf=\rho(\cos\varphi_0 \hat{\bf x}+\sin\varphi_0 \hat{\bf y})$ and recalling (\ref{eq:kappan}), whence \cite{Watson}
\begin{eqnarray}
\left<
\exp\left(i\omega
\kappabf_n\cdot {\rhobf}
\right)
\right>&=&\frac{1}{2\pi}\int\limits_{-\pi}^{\pi}\exp\left[
\frac{i\omega\rho\cos\left(\phi_n-\varphi_0\right)}{c_0}
\right]d\phi_n\nonumber\\
&=&J_0\left(\frac{\omega \rho}{c_0}\right).
\end{eqnarray}
Finally, recalling (\ref{eq:gamma}), we obtain
\beq
c(\rho,\tau)\sim
\left(
\frac{\gamma}{4 \pi T_s}\right)
\int\limits_{-\infty}^\infty  
|S(\omega)|^2
J_0\!\left(\frac{\omega \rho}{c_0}\right)
\exp\left(-i \omega \tau\right)
d\omega,
\label{eq:ce3}
\eeq
which is structurally identical, apart from the multiplicative constant, to the result in (\ref{eq:ce2}) obtained via a heuristic argument. From (\ref{eq:ce2}) and (\ref{eq:ce3}), Eq. (\ref{eq:FP}) readily follows.


\begin{thebibliography}{99}


\bibitem{Lichtenberg}{A. J. Lichtenberg and M. A. Liebermann, {\em Regular and Stochastic Motion}. New York (NY):
Springer-Verlag, 1983.}

\bibitem{Ott}{E. Ott, {\em Chaos in Dynamical Systems}. Cambridge (UK): Cambridge University Press, 1993.}

\bibitem{Born}{M. Born, E. Wolf, and A. B. Bhatia, {\em Principles of Optics: Electromagnetic Theory of Propagation, Interference and Diffraction of Light}. Cambridge (UK): Cambridge University Press, 1999.}

\bibitem{Sinai}{Ya. G. Sinai, ``Dynamical systems with elastic reflections,'' {\em Russ. Math. Surveys}, vol. 25, No. 2, pp. 137--189, 1970.}

\bibitem{Bunimovich}{L. A. Bunimovich, ``Conditions of stochasticity of two-dimensional billiards,'' {\em Chaos}, vol. 1, No. 2, pp. 187--193, Aug. 1991.}

\bibitem{Wirzba}{A. Wirzba, ``Quantum mechanics and semiclassics
of hyperbolic $n$-disk scattering systems,'' {\em Phys. Rep.}, vol. 309, No. 1-2, pp. 1--116, Feb. 1999.}

\bibitem{Gaspard}{T. Harayama and P. Gaspard, ``Diffusion of particles bouncing on a one-dimensional periodically corrugated floor,'' {\em Phys. Rev. E}, vol. 64, 036215, Aug. 2001.}

\bibitem{Chernov}{N. Chernov and R. Markarian, {\em Chaotic Billiards}. Providence (RI): American Mathematical Society, 2006.}

\bibitem{Smilansky}{T. Kottos, U. Smilansky, J. Fortuny, and G. Nesti, ``Chaotic scattering of microwaves,'' {\em Radio Sci.}, vol. 34, No. 4, pp. 747--758, July-Aug. 1999.}

\bibitem{Mackay1}{A. J. Mackay, ``Application of chaos theory to ray tracing in ducts,'' {\em IEE Proc. Radar Sonar Navig.}, vol. 164, No. 6, pp. 298--304, Dec. 1999.}

\bibitem{Anlage1}{S. Hemmady, X. Zheng, E. Ott, T. M. Antonsen, and S. M. Anlage, ``Universal impedance fluctuations in wave chaotic systems,'' {\em Phys. Rev. Lett.}, vol. 94, No. 1, 014102, Jan. 2005.}

\bibitem{APMag}{V. Galdi, I. M. Pinto, and L. B. Felsen, ``Wave propagation in ray-chaotic enclosures: Paradigms, oddities and examples,'' {\em IEEE Antennas Propagat. Mag.}, vol. 47, No. 1, pp. 62--81, Feb. 2005.}

\bibitem{APchaos}{G. Castaldi, V. Fiumara, V. Galdi, V. Pierro, I. M. Pinto, and L. B. Felsen, ``Ray-chaotic footprints in deterministic wave dynamics: A test model with coupled Floquet-type and ducted-type mode characteristics,'' {\em IEEE Trans. Antennas Propagat.}, vol. 53, No. 2, pp. 753--765, Feb. 2005.}

\bibitem{Anlage2}{S. Hemmady, X. Zheng, T. M. Antonsen, E. Ott, and S. M. Anlage, ``Universal statistics of the scattering coefficient of chaotic microwave cavities,'' {\em Phys. Rev. E}, vol. 71, No. 5, 056215, May 2005.}

\bibitem{Anlage3}{X. Zheng, S. Hemmady, T. M. Antonsen, S. M. Anlage, and E. Ott, ``Characterization of fluctuations of impedance and scattering matrices in wave chaotic scattering,'' {\em Phys. Rev. E}, vol. 73, No. 4, 046208, Apr. 2006.}

\bibitem{APchaos_cyl}{G. Castaldi, V. Galdi, and I. M. Pinto, ``A study of ray-chaotic cylindrical scatterers,'' {\em IEEE Trans. Antennas Propagat.}, vol. 56, No. 8, pp. 2638--2648, Aug. 2008.}

\bibitem{Ramahi}{F. Seydou, T. Seppanen, and O. M. Ramahi, ``Computation of the Helmholtz eigenvalues in a class of chaotic cavities using the multipole expansion technique,'' {\em IEEE Trans. Antenans. Propagat.}, vol. 57, No. 4, Part 2, pp. 1169--1177, Apr. 2009.}

\bibitem{Noeckel2}{J. U. N\"ockel and A. D. Stone, ``Ray and wave chaos in asymmetric resonant optical cavities,''
{\em Nature}, vol. 385, No. 6611, pp. 45--47, Jan. 1997.}

\bibitem{Gmachl}{C. Gmachl, F. Capasso, E. E. Narimanov, J. U. N\"ockel, A. D. Stone, J. Faist, D. L. Sivco, and A. Y. Cho, ``High-power directional emission from microlasers with chaotic resonators,'' {\em Science}, vol. 280, No. 5369, pp. 1556--1564, June 1998.}

\bibitem{Gensty}{T. Gensty, K. Becker, I. Fischer, W. Els\"a\ss{}er, C. Degen, P. Debernardi, and G. P. Bava, ``Wave chaos in real-world vertical-cavity surface-emitting lasers,'' {\em Phys. Rev. Lett.}, vol. 94, No. 23, 233901, June 2005.}

\bibitem{Shinohara}{S. Shinohara, T. Harayama, T. Fukushima, M. Hentschel, T. Sasaki, and E. E. Narimanov,
``Chaos-assisted directional light emission from microcavity lasers,'' {\em Phys. Rev. Lett.}, vol. 104, No. 16, 163902, Apr. 2010.}



\bibitem{Mackay2}{A. Mackay, ``Random wave methods for the prediction of the RCS of homogeneous chaotic straight ducts,'' {\em IEE Proc. Radar Sonar Navig.}, vol. 148, No. 6, pp. 331--337, Dec. 2001.}

\bibitem{Mackay3}{A. Mackay, ``Random wave and Schell model for the mean RCS of bent chaotic ducts with a homogeneous scattered aperture field distribution,'' {\em IEE Proc. Radar Sonar Navig.}, vol. 148, No. 6, pp. 338--342, Dec. 2001.} 



\bibitem{Fink1}{C. Draeger and M. Fink, ``One-channel time reversal of elastic waves in a chaotic 2D-silicon cavity,''
{\em Phys. Rev. Lett.}, vol. 79, No. 3, pp. 407--410, July 1997.}

\bibitem{Anlage4}{S. M. Anlage, J. Rodgers, S. Hemmady, J. Hart, T. M. Antonsen, and E. Ott, ``New results in chaotic time-reversed electromagnetics: High frequency one-recording-channel time reversal mirror,'' {\em Acta Physica Polonica A}, vol. 112, No. 4, pp. 569--574, Oct. 2007.} 

\bibitem{Anlage5}{B. T. Taddese, J. Hart, T. M. Antonsen, E. Ott, and S. M. Anlage, ``Sensor based on extending the concept of fidelity to classical waves,'' {\em Appl. Phys. Lett.}, vol. 95, No. 11, 114103, Sep. 2009.}

\bibitem{Cappetta}{L. Cappetta, M. Feo, V. Fiumara, V. Pierro, and I. M. Pinto,  ``Electromagnetic chaos in mode stirred reverberation enclosures,'' {\em IEEE Trans. Electromagn. Compatibility}, vol. 40, No. 3, pp. 185--192, Aug. 1998.}

\bibitem{Orjubin1}{G. Orjubin, E. Richalot, O. Picon, and  O. Legrand, ``Chaoticity of a reverberation chamber assessed from the analysis of modal distributions obtained by FEM,''  {\em IEEE Trans. Electromagnetic Compatibility},
vol. 49,  No. 4, pp. 762--771, Nov. 2007.}

\bibitem{Orjubin2}{G. Orjubin, E. Richalot, O. Picon, and O. Legrand, ``Wave chaos techniques to analyze a modeled reverberation chamber,'' {\em Comptes Rendus Physique}, vol. 10, No. 1, pp. 42--53, Jan. 2009.}



\bibitem{Berry}{M. V. Berry, ``Quantum chaology,'' {\em Proc. Roy. Soc. London}, vol. A413, No. 1844, pp. 183--198, Sep. 1987.}

\bibitem{Gutz}{M. C. Gutzwiller,  {\em Chaos in Classical and Quantum Mechanics}. New York (NY): Springer-Verlag, 1990.}

\bibitem{Haake}{F. Haake, {\em Quantum Signatures of Chaos}. New York (NY): Springer-Verlag, 1991.}

\bibitem{Stockmann}{H.-J. St\"ockmann, {\em Quantum Chaos. An Introduction}. Cambridge (UK): Cambridge University Press, 1999.}


\bibitem{RM}{M. L. Mehta, {\em Random Matrices}, 2nd ed. San Diego (CA): Academic Press, 1991.}

\bibitem{Heller1}{E. J. Heller, ``Bound-state eigenfunctions of classically chaotic Hamiltonian systems: Scars of periodic orbits,'' {\em Phys. Rev. Lett.}, vol. 53, No. 16, pp. 1515--1518, Oct. 1984.}

\bibitem{Sridhar}{S. Sridhar, ``Experimental observation of scarred eigenfunctions of chaotic microwave cavities,''
{\em Phys. Rev. Lett.}, vol. 67, No. 7, pp. 785--788, Aug. 1991.}

\bibitem{Berry1}{M. V. Berry, ``Regular and irregular semiclassical wavefunctions,'' {\em J. Phys. A: Math. Gen.}, vol. 10, No. 12, pp. 2083--2091, Dec. 1977.}

\bibitem{Heller2}{P. O'Connor, J. Gehlen, and E. J. Heller, ``Properties of random superpositions of plane waves,'' {\em Phys. Rev. Lett.}, vol. 58, No. 13, pp. 1296--1299, Mar. 1987.}
 
\bibitem{Hill}{D. A. Hill, ``Plane wave integral representation for fields in reverberation chambers,'' {\em IEEE Trans. Electromag. Compat.}, vol. 40, No. 3, pp. 209--217, Aug. 1998.}
  
\bibitem{Gorin}{T. Gorin, T. Prosen, T. H. Seligman, and M \v{Z}nidari\v{c}, ``Dynamics of Loschmidt echoes and fidelity decay,'' {\em  Phys. Rep.}, vol. 435, No. 2-5, pp. 33--156, Nov. 2006.}
 
\bibitem{Antonsen}{J. A. Hart, T. M. Antonsen, and E. Ott, ``Scattering a pulse from a chaotic cavity: Transitioning from algebraic to exponential decay,'' {\em Phys. Rev. E}, vol. 79, No. 1, 016208, Jan. 2009.}

\bibitem{ICEAA99}{V. Fiumara, V. Galdi, V. Pierro, and I. M. Pinto, ``From mode-stirred enclosures to electromagnetic Sinai billiards: Chaotic models of reverberation enclosures,'' Proc. 6th Int. Conference on Electromagnetics in Advanced Applications (ICEAA), Torino, Italy, Sep. 15-17, 1999, pp. 357--360.}

\bibitem{Taflove}{A. Taflove and S. C. Hagness, {\em Computational Electrodynamics: The Finite-Difference Time-Domain Method}, 3rd Ed. Norwood (MA): Artech-House, 2005.}

\bibitem{Kudrolli}{A. Kudrolli, S. Sridhar, A. Pandey, and R. Ramaswamy, ``Signatures of chaos in quantum billiards: Microwave experiments,'' {\em Phys. Rev. E}, vol. 49, No. 1, pp. R11--R14, Jan. 1994.}

\bibitem{Stratton}{J. A. Stratton, {\em Electromagnetic Theory}. Piscataway (NJ): Wiley-IEEE Press, 2007.}

\bibitem{Berry2}{M. V. Berry, J. P. Keating, and H. Schomerus, ``Universal twinkling exponents for spetral fluctuations associated with mixed chaology,'' {\em Proc. R. Soc. Lond. A}, vol. 456, No. 1999, pp. 1659--1668, July 2000.}

\bibitem{Backer}{A. B\"acker and R. Schubert, ``Autocorrelation function of eigenstates in chaotic and mixed systems,'' {\em J. Phys. A: Math. Gen.}, vol. 35, No. 3, pp. 539--564, Jan. 2002.} 

\bibitem{Middleton}{D. Middleton, {\em An Introduction to Statistical Communication Theory}. Piscataway (NJ): IEEE Press, 1996.}

\bibitem{Le}{N. D. Le and J. V. Zidek, {\em Statistical Analysis of Environmental Space-Time Processes}. New York (NY): Springer, 2006.}

\bibitem{Watson}{G. N. Watson, {\em A Treatise on the Theory of Bessel Functions}, 2nd Ed. Cambridge (UK), Cambridge University Press, 1995.}

\bibitem{Heyman}{E. Heyman and L. B. Felsen, ``Weakly dispersive spectral theory of transients, part I: Formulation and interpretation,'' {\em IEEE Trans. Antennas Propagat.}, vol.35, No. 1, pp. 80--86, Jan. 1987.}

\bibitem{Prudnikov}{A. P. Prudnikov, Yu. A. Brychkov, and O. I. Marichev, {\em Integral and Series: Special Functions}, Vol. 2. New York (NY): Gordon \& Breach, 1986.}

\bibitem{Abramowitz}{M. Abramowitz and I. E. Stegun, {\em Handbook of Mathematical Functions}. New York (NY): Dover, 1972.}

\bibitem{KS}{R. Bartoszy\'nski, M. Niewiadomska-Bugaj, {\em Probability and Statistical Inference}, 2nd Ed. New York (NY): Wiley, 2008.}

\bibitem{Wald}{A. Wald, ``On cumulative sums of random variables,'' {\em Ann. Math. Stat.}, vol. 15, No. 3, pp. 283--296, Sep. 1944.}

\end{thebibliography}
\end{document}